\documentclass[lettersize,journal]{IEEEtran}
\IEEEoverridecommandlockouts
\usepackage{amsmath,amsfonts,amsthm,amssymb}
\usepackage{algorithmic}
\usepackage{array}
\usepackage{textcomp}
\usepackage{stfloats}
\usepackage{url}
\usepackage{cite}
\usepackage{verbatim}
\usepackage{graphicx}
\usepackage{hyperref}
\usepackage{subcaption, multirow, makecell}
\usepackage{mathtools}
\usepackage{float}
\DeclarePairedDelimiter\ceil{\lceil}{\rceil}

\newtheorem{thm}{Theorem}

\newtheorem{example}{Example}

\newtheorem{rem}{Remark}
\def\BibTeX{{\rm B\kern-.05em{\sc i\kern-.025em b}\kern-.08em
    T\kern-.1667em\lower.7ex\hbox{E}\kern-.125emX}}
\usepackage{balance}
\begin{document}
\title{A Novel Coded Caching Scheme for Partially Cooperative Device-to-Device Networks}
\author{Rashid Ummer N.T.\IEEEauthorrefmark{1}, \textit{Member, IEEE}, \IEEEauthorblockN{K. K. Krishnan Namboodiri\IEEEauthorrefmark{2}, \textit{Member, IEEE}, and B. Sundar Rajan\IEEEauthorrefmark{1}, \textit{Life Fellow, IEEE}} 
\thanks{\IEEEauthorblockA{\IEEEauthorrefmark{1} Department of Electrical Communication Engineering, Indian Institute of Science, Bengaluru, India  (e-mail: \{rashidummer, bsrajan\}@iisc.ac.in).}} \\ \thanks{\IEEEauthorblockA{\IEEEauthorrefmark{2} Communication System Department, Eurecom, Biot, France (e-mail: krishnan.karakkad@eurecom.fr).}}}


\maketitle

\begin{abstract}
Device-to-device (D2D) communication is one of the most promising techniques for future wireless cellular communication systems. This paper considers coded caching in a partially cooperative wireless D2D network, where only a subset of users transmit during delivery, while all users request files. The non-transmitting users are referred to as selfish users. All existing schemes that do not require knowledge of the identity of selfish users before content placement are limited to the high-memory regime, particularly when the number of selfish users is large. We propose a novel coded caching scheme for a partially cooperative D2D network that operates in all feasible memory regimes, regardless of the number of selfish users. We also derive a lower bound on the transmission load of a partially cooperative D2D coded caching scheme. Using this bound, the proposed scheme is shown to be optimal in the high-memory regime.
\end{abstract}

\begin{IEEEkeywords}
Coded caching, D2D network, partially cooperative, selfish users.
\end{IEEEkeywords}

\section{Introduction}

Device-to-device (D2D) communication plays an important role in realizing the Internet of Things (IoT) in future fifth-generation (5G) and beyond wireless cellular communication systems \cite{TUY, X, LLLW}. A survey on advantages, challenges, and solutions of using D2D communication in 5G was carried out in  \cite{ARY}. D2D communication is also a key emerging technique for next-generation wireless IoT networks \cite{MACF, BZ, ERW}. Edge caching assisted device-to-device (D2D) communication has been recognized as a promising solution for reducing network latency and backhaul transmission load \cite{FSYWQ, ZTS}. Cache-aided D2D network has been studied extensively in the literature due to its significance in cellular communication systems and IoT networks \cite{GJMDC, GMDC, JCM2, ZWSW, JCM3, CY, ZFLLX, LZC}. Coded caching in a wireless D2D network was first studied by Ji, Caire, and Molisch in \cite{Ji} (referred to as the JCM scheme), motivated by the coded caching technique introduced by Maddah-Ali and Niesen in \cite{MaN} for a broadcast network (referred to as the MAN scheme). A D2D coded caching scheme operates in two phases: a \textit{placement phase} followed by a \textit{delivery phase}. In the placement phase, a central server places content in all user caches. In the delivery phase, the central server is not present, and the demands of all users are served through inter-user coded multicast transmissions. The transmission load of the JCM scheme is shown to be order optimal in \cite{Ji} and \cite{CKRG}. An improved lower bound for the transmission load of a D2D coded caching scheme was derived in \cite{ST}.

Various aspects of coded caching in a D2D network have been studied in \cite{WCJ_RS, WCJ_HY, PBHT, WCYT, IZY, ZY, ZYJ, LC, RS}. The JCM scheme and all other schemes in \cite{WCJ_RS, WCJ_HY, PBHT, WCYT, IZY, ZY, ZYJ, LC, RS} assume by default that all users in the D2D network do data transmission during the delivery phase. However, in practical D2D communication scenarios, it is highly probable that many of the users show selfish behaviors \cite{GZCLJC}. A user who does not transmit data in the delivery phase is referred to as a \textit{selfish user}. The selfish behavior can be for saving energy or due to privacy and security concerns \cite{GZCLJC}. Even though selfish users do not transmit data, they can request files. A D2D network that consists of selfish users is referred to as a \textit{partially cooperative D2D network}. Coded caching in a partially cooperative D2D network was first studied by Tebbi and Sung in \cite{TS}. A $(K, S, N)$ partially cooperative D2D network consists of $K$ users, of which $S$ are selfish, each equipped with a cache of size $M$ files, and a central server with a library of $N$ files. 

In \cite{TS}, the authors proposed two partially cooperative D2D coded caching schemes, a deterministic caching scheme and a random caching scheme. In the deterministic caching scheme, each user caches a predetermined subfiles of each file. Whereas, in the random caching scheme, each user randomly caches a set of coded subfiles of each file. The authors in \cite{PV} proposed a partially cooperative D2D coded caching scheme, which has an improved transmission load compared to the deterministic caching scheme in \cite{TS}. Guan \textit{et al.} in \cite{GHXZCH} proposed three partially cooperative D2D coded caching schemes, namely Scheme A, Scheme B, and Scheme C. Scheme A in \cite{GHXZCH} outperforms the deterministic scheme in \cite{TS} and the scheme in \cite{PV}. However, the deterministic scheme in \cite{TS}, the scheme in \cite{PV} and the Scheme A in \cite{GHXZCH}  operate in higher memory regimes, that is when $M \ge \frac{N}{K}(S+1)$, and also need to know the identity of all the selfish users to start the delivery phase. Scheme B and Scheme C in \cite{GHXZCH} require prior knowledge of the identity of selfish users before the placement phase itself. A lower bound on the transmission load was also obtained in \cite{GHXZCH}, but for the limiting case of $S=1$.  We propose a partially cooperative D2D coded caching scheme that operates in all feasible memory regimes, regardless of the number of selfish users. 

In contrast to existing schemes, the design of transmissions in our scheme does not rely on the identities of selfish users; each non-selfish user can generate its transmissions independently of the others. This independence ensures that the delivery scheme does not need to be redesigned depending on which users are selfish, and also reduces control or coordination overhead during delivery. For successful decoding, however, users must be aware of the set of non-selfish users. We also obtain a general lower bound on the transmission load of a partially cooperative D2D coded caching scheme. The contributions in this paper are summarized as follows: \\
\noindent $\bullet$ A novel coded caching scheme for a partially cooperative D2D network is proposed. This scheme operates in all feasible memory regimes, irrespective of the number of selfish users. Compared to the deterministic scheme in \cite{TS}, the scheme in \cite{PV}, and Scheme A in \cite{GHXZCH}, the proposed scheme additionally operates in the memory regime $\frac{N}{K-S} \le M \le \frac{N}{K}(S+1)$, which is not covered by those schemes. Unlike existing schemes, the proposed approach does not require knowledge of the identities of selfish users to design either the placement phase or the delivery phase; it only requires knowledge of the maximum possible number of selfish users in the D2D network.  \\ 
\noindent $\bullet$ A general cut-set based lower bound on the transmission load of a partially cooperative D2D coded caching scheme is obtained. Using this bound, the proposed scheme is shown to be optimal in the memory regime $M\geq \frac{N}{1+\frac{K-S-1}{(K-1)^2}}$.

The rest of the paper is organized as follows. In Section \ref{prelim_pcd2d}, we describe the system model. The proposed scheme and the proposed lower bound are discussed in Section \ref{scheme_pcd2d}. In Section \ref{perform_anlysis_pcd2d}, the performance analysis of the proposed scheme is carried out. Section \ref{concl_pcd2d} concludes the paper.  

\textit{Notations}: For any positive integer $n$ and any integer $m <n$, $[n]$ denotes the set $\{1,2,...,n\}$ and $[m:n]$ denotes the set $\{m,m+1,...,n\}$. For integers $n$ and $i$, the binomial coefficient $\binom{n}{i}$ is defined as $\frac{n!}{i!(n-i)!}$ when $0 \le i \le n$, and $0$ when $i <0$. For a set $\mathcal{A}$ and a positive integer $i \leq |\mathcal{A}|$,  $\binom{\mathcal{A}}{i}$ denotes all the $i$-sized subsets of $\mathcal{A}$. For sets $\mathcal{A} \text{ and }\mathcal{B}$, $\mathcal{A} \backslash \mathcal{B}$ denotes the elements in $\mathcal{A}$ but not in $\mathcal{B}$.

\section{System Model}\label{prelim_pcd2d}
\begin{figure}[!htbp]
	\centering
	\captionsetup{justification=centering}
	\includegraphics[width=0.49\textwidth]{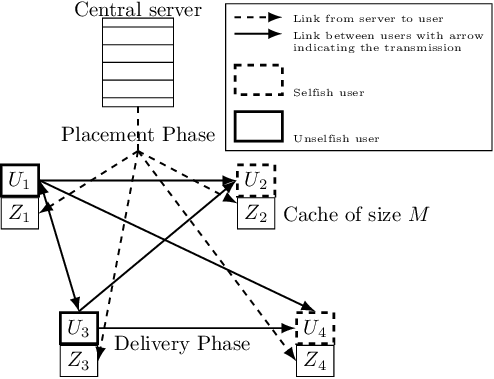}
	\caption{Partially cooperative D2D coded caching network model with $K=4$ and $S=2$.}
	\label{fig_pcd2d}
\end{figure}

In this section, we review the partially cooperative D2D network model. A $(K, S, N)$ partially cooperative D2D network consists of a central server with a library of $N$ files $\{W_n: n \in [N]\}$ each of size $B$ symbols over a finite field $\mathbb{F}_q$  \footnote{We assume that $q$ is large enough such that all MDS codes considered in this work exist over $\mathbb{F}_q$}, and $K \le N$ users $\{U_k: k \in [K]\}$ each equipped with a cache of size $M$ files, among which $S$ users are selfish. The condition $M\geq \frac{N}{K-S}$ is needed to ensure that any possible demands can be met using the cache contents of the $(K-S)$ non-selfish users. An illustration of the network model for $K=4$ and $S=2$ is shown in Fig.\ref{fig_pcd2d}. 

The partially cooperative D2D coded caching scheme operates in two phases. \\
\noindent $\bullet$ \textbf{Placement phase}: During this phase, each file $W_{n}$ is split into $F$ non-overlapping subfiles (referred to as subpacketization level), i.e., $W_n =\{W_{n, i}: i \in [F]\}$. Each user $U_k$, where $k \in [K]$, caches $MB$ symbols in its cache, denoted by $Z_k$. The cache placement can be coded or uncoded. During the placement phase, the later demands of the users are unknown. We assume that only the number of selfish users, and not their identities, is known before the placement phase. \\
\noindent $\bullet$ \textbf{Delivery phase}:  During this phase, each user requests a file from $\{W_n: n \in [N]\}$, with the set of requested files represented by a demand vector $\vec{d}=(d_1,\ldots,d_K)$. Even though all users request a file, only the $(K-S)$ non-selfish users do the data transmission. Let $\mathcal{U}$ and $\mathcal{U_S}$ denote the set of users and the set of selfish users, respectively. For a given $\vec{d}$, each non-selfish user $U_k \in \mathcal{U} \backslash \mathcal{U_S}$ broadcasts a coded message $X_{k,\vec{d}}$ consisting of $S_{k,\vec{d}}$ subfiles using its cache content. The delivery is assumed to be \textit{one-shot}, meaning that each user can recover any bit of its requested file using its own cache content and at most one transmission from other users. Using these transmissions along with its cache content, each user can retrieve its requested file. The corresponding worst-case transmission load $R$ normalized to the file size is given by {\small$ R \triangleq \max_{\vec{d}} \left(\sum_{k : U_k \in \mathcal{U} \backslash \mathcal{U_S}} S_{k,\vec{d}}\right) / F $}.

\section{Proposed partially cooperative D2D coded caching scheme}\label{scheme_pcd2d}
In this section, we first present the novel partially cooperative D2D coded caching scheme. Then, a lower bound on the transmission load is derived to show the optimality of the proposed scheme in a high memory regime. The following theorem states the achievable load-memory pairs of the proposed scheme.
\begin{thm}\label{thm_pcd2d}
	For the \((K, S, N)\) partially cooperative D2D coded caching network, the following memory-load pair is achievable, 
	{\small\begin{equation}\label{eq:thm_pcd2d}
		\begin{split}
		(M,R)= & \left(\frac{N(t+1)(K-1)}{(K-S)(K-1)+tS}, \right. \\ & \left. \frac{(K-S)(K-1)}{(K-S)(K-1)+tS}\frac{K-t-1}{t+1}\right),
		\end{split}
	\end{equation}}
	for every \(t\in [0:K-1]\). 
\end{thm}
\begin{IEEEproof}
	The partially cooperative D2D coded caching scheme that achieves (\ref{eq:thm_pcd2d}) is described below.\\
		\noindent $\bullet$ \textbf{Placement Phase :}
		Let \(t\in [0:K-1]\). Each file \(W_n,n\in [N]\) is divided into \((K-S)\binom{K-1}{t}+S\binom{K-2}{t-1}\) non-overlapping subfiles, i.e.,
		{\small\[W_n = \left\{W_{n,i}:i\in \left[(K-S)\binom{K-1}{t}+S\binom{K-2}{t-1}\right]\right\}.\]}
		These subfiles are then encoded using Maximum Distance Separable (MDS) codes. Note that, an $[n,k]$ MDS code has the property that the $k$ information symbols can be reconstructed from any subset of $k$ code symbols out of the $n$ symbols. Let \(\mathbf{G}\) be a generator matrix of a \(\left[K\binom{K-1}{t},(K-S)\binom{K-1}{t}+S\binom{K-2}{t-1}\right]\) MDS code. The subfiles of \(W_n, \forall n\in [N]\) are then encoded using \(\mathbf{G}\) as, 
		{\small\begin{equation*}
			\begin{split}
				\left[Y_{n,j}:j\in  \left[K\binom{K-1}{t}\right]\right] = \\ & \hspace{-3cm} \mathbf{G}\left[W_{n,i}:i\in \left[(K-S)\binom{K-1}{t}  +  S\binom{K-2}{t-1}\right]\right]. 
			\end{split}
		\end{equation*}}
		Every \(j\in [K\binom{K-1}{t}]\) can be uniquely represented as 
		$j=(k-1)\binom{K-1}{t}+\phi_k(\mathcal{T})$ 
		for some \(k\in [K]\) and \(\mathcal{T}\subseteq [K]\backslash\{k\}\) with \(|\mathcal{T}|=t\), where the function \(\phi_k:\binom{[K]\backslash \{k\}}{t}\rightarrow \left[\binom{K-1}{t}\right]\) maps a subset \(\mathcal{T}\) to its lexicographic index in \(\binom{[K]\backslash \{k\}}{t}\). Then, for every \(n\in [N]\) and \(j\in [K\binom{K-1}{t}]\), we define
		\[Y_{n,\mathcal{T}}^{(k)}\triangleq Y_{n,j}\]
		where \(j=(k-1)\binom{K-1}{t}+\phi_k(\mathcal{T})\).
		For every \(k\in [K]\), the content stored in the cache of user $U_k$, denoted by  $Z_k$, is given by 
		{\small\begin{align*}
			Z_k &= \left\{Y_{n,\mathcal{T}}^{(k)}: \mathcal{T}\subseteq[K]\backslash \{k\},|\mathcal{T}|=t,n\in [N]\right\}\cup\\& \hspace{-2cm} \qquad\qquad\left(\bigcup_{\ell \in [K]\backslash \{k\}}\left\{Y_{n,\mathcal{T}}^{(\ell)}: \mathcal{T}\ni k, \mathcal{T}\subseteq[K]\backslash \{\ell\},|\mathcal{T}|=t,n\in [N]\right\}\right).
		\end{align*}}
		Note that, the cache size 
		{\small\begin{align*}
			M &=|Z_k|  = \frac{N\binom{K-1}{t}}{(K-S)\binom{K-1}{t}+S\binom{K-2}{t-1}}+ \\&  \frac{N(K-1)\binom{K-2}{t-1}}{(K-S)\binom{K-1}{t}+S\binom{K-2}{t-1}}
			 = \frac{N(t+1)(K-1)}{(K-S)(K-1)+tS}.
		\end{align*}}
		
		\noindent $\bullet$ \textbf{Delivery Phase:}
		Suppose user $U_s$ requests for the file \(W_{d_s}\), where \(s\in [K]\) and \(d_s\in [N]\). Let \(\mathcal{U_K}: \mathcal{K}\subseteq [K]\) with \(|\mathcal{K}|=(K-S)\) be the set of transmitting users/non-selfish users. User $U_k$, \(k\in\mathcal{K}\) makes coded transmissions corresponding to every set \(\mathcal{S}\subseteq [K]\backslash\{k\}\) with \(|\mathcal{S}|=t+1\). The transmission by user $U_k$, \(k\in \mathcal{K}\), corresponding to some \(\mathcal{S}\subseteq [K]\backslash \{k\}\) with \(|\mathcal{S}|=t+1\) is
		\begin{equation*}
			\bigoplus_{s\in \mathcal{S}} Y_{d_s,\mathcal{S}\backslash \{s\}}^{(k)}.
		\end{equation*}
		Note that, user \(U_k\) has access to \(Y_{n,\mathcal{T}}^{(k)}\) for every \(n\in [N]\) and \(\mathcal{T}\in \binom{[K]\backslash\{k\}}{t}\), and thus it can create the transmitted coded message. Consequently, the load incurred is 
		{\small\begin{align*}
			R&=\frac{(K-S)\binom{K-1}{t+1}}{(K-S)\binom{K-1}{t}+S\binom{K-2}{t-1}}\\
			&=\frac{(K-S)(K-1)}{(K-S)(K-1)+tS}\frac{K-t-1}{t+1}.
		\end{align*}}
		
		\noindent $\bullet$ \textbf{Decodability:}
		From the property of MDS codes, user \(U_k\) can decode its demanded file \(W_{d_k}\) from any distinct \(Q = (K-S)\binom{K-1}{t}+S\binom{K-2}{t-1}\) coded subfiles of it. In other words, using any \(Q\) coded subfiles from the set \(\{Y_{d_k,\mathcal{T}}^{(\ell)}: \ell\in [K],\mathcal{T}\subseteq [K]\backslash \{\ell\},|\mathcal{T}|=t\}\), the file \(W_{d_k}\) can be decoded. Note that, for every \(n\in [N]\) and \(k\in [K]\), we have
		$
			|\left\{Y_{n,\mathcal{T}}^{(k)}: \mathcal{T}\subseteq[K]\backslash \{k\},|\mathcal{T}|=t\right\}|=\binom{K-1}{t},
		$
		and for every \(\ell\in [K]\backslash \{k\}\), we have
		$
			|\left\{Y_{n,\mathcal{T}}^{(\ell)}: \mathcal{T}\ni k, \mathcal{T}\subseteq[K]\backslash \{\ell\},|\mathcal{T}|=t\right\}|=\binom{K-2}{t-1}.
		$
		Therefore, each user has access to \(Q_z = \binom{K-1}{t}+(K-1)\binom{K-2}{t-1}\) coded subfiles of every file from their caches. 
		We now consider the decodability of selfish users and non-selfish users separately. 
		
		First, let user \(U_k\), \(k\in\mathcal{K}\), be a non-selfish user. Let \(k'\in \mathcal{K}\) such that \(k'\neq k\). Consider a set \(\mathcal{T}\subseteq [K]\backslash \{k,k'\}\) such that \(|\mathcal{T}|=t\), and define the set \(\mathcal{S} = \mathcal{T}\cup\{k\}\). Therefore, we have \(|\mathcal{S}|=t+1\). Consider the following transmission from user \(U_{k'}\), corresponding to the set \(\mathcal{S}\)
		{\small\begin{align*}
			\bigoplus_{s\in \mathcal{S}} Y_{d_s,\mathcal{S}\backslash \{s\}}^{(k')}=Y_{d_k,\mathcal{T}}^{(k')}\oplus \left(\bigoplus_{s\in \mathcal{S},s\neq k} Y_{d_s,\mathcal{S}\backslash \{s\}}^{(k')}\right).
		\end{align*}}
		Since \(k\in \mathcal{S}\backslash \{s\}\) for every \(s\neq k\), user \(U_k\) has the coded subfile \(Y_{d_s,\mathcal{S}\backslash \{s\}}^{(k')}\), for every \(k'\in\mathcal{K}\), in its cache. Therefore, from the delivery phase, user \(U_k\) can decode \(Y_{d_k,\mathcal{T}}^{(k')}\) for every \(k'\in \mathcal{K}\backslash\{k\}\) and \(\mathcal{T}\subseteq [K]\backslash\{k,k'\}\). For a \(k,k'\in \mathcal{K}\), we have
		 $
			\left|\left\{Y_{d_k,\mathcal{T}}^{(k')}:\mathcal{T}\subseteq [K]\backslash\{k,k'\}\right\}\right|=\binom{K-2}{t}.
		$
		Thus, in addition to the subfiles in the cache, user \(U_k\) can get \(Q_T = (K-S-1)\binom{K-2}{t-1}\) coded subfiles of \(W_{d_k}\) from the delivery phase. Therefore, user \(U_k\) has a total of
		{\small
			\begingroup
			\allowdisplaybreaks
		\begin{align*}
			&Q_z+Q_T\notag\\&=\binom{K-1}{t}+(K-1)\binom{K-2}{t-1}+(K-S-1)\binom{K-2}{t}\\&=\binom{K-1}{t} +S\binom{K-2}{t-1}\\&+(K-S-1)\left(\binom{K-2}{t-1}+\binom{K-2}{t}\right)\\&=\binom{K-1}{t}+S\binom{K-2}{t-1}+(K-S-1)\binom{K-1}{t}\\&=(K-S)\binom{K-1}{t}+S\binom{K-2}{t-1}=Q
		\end{align*}
	\endgroup}
		coded subfiles of \(W_{d_k}\). Thus, the user \(U_k\)  can decode \(W_{d_k}\).
		
		Let user \(U_k\), \(k\in [K]\backslash \mathcal{K}\), be a selfish user. Consider user \(U_{k'}\) such that \(k'\in \mathcal{K}\).
		Consider a set \(\mathcal{T}\subseteq [K]\backslash \{k,k'\}\) such that \(|\mathcal{T}|=t\), and define the set \(\mathcal{S} = \mathcal{T}\cup\{k\}\). Therefore, we have \(|\mathcal{S}|=t+1\). Consider the following transmission from user \(U_{k'}\), corresponding to the set \(\mathcal{S}:\)
		{\small\begin{align*}
			\bigoplus_{s\in \mathcal{S}} Y_{d_s,\mathcal{S}\backslash \{s\}}^{(k')}=Y_{d_k,\mathcal{T}}^{(k')}\oplus \left(\bigoplus_{s\in \mathcal{S},s\neq k} Y_{d_s,\mathcal{S}\backslash \{s\}}^{(k')}\right).
		\end{align*}}
		Since \(k\in \mathcal{S}\backslash \{s\}\) for every \(s\neq k\), user \(U_k\) has the coded subfile \(Y_{d_s,\mathcal{S}\backslash \{s\}}^{(k')}\), for every \(k'\in\mathcal{K}\), in its cache. Therefore, from the delivery phase, user \(U_k\) can decode \(Y_{d_k,\mathcal{T}}^{(k')}\) for every \(k'\in \mathcal{K}\) and \(\mathcal{T}\subseteq [K]\backslash\{k,k'\}\). For a \(k\in [K]\backslash\mathcal{K}\) and \(k'\in \mathcal{K}\), we have
		$
			\left|\left\{Y_{d_k,\mathcal{T}}^{(k')}:\mathcal{T}\subseteq [K]\backslash\{k,k'\}\right\}\right|=\binom{K-2}{t}.
		$
		Thus, in addition to the subfiles in the cache, user \(U_k\) can get \(Q_T = (K-S)\binom{K-2}{t-1}\) coded subfiles of \(W_{d_k}\) from the delivery phase. Therefore, user \(U_k\) has a total of
		{\small
			\begingroup
			\allowdisplaybreaks
		\begin{align}
			&Q_z+Q_T\notag\\&=\binom{K-1}{t}+(K-1)\binom{K-2}{t-1}+(K-S)\binom{K-2}{t}\notag\\&>\binom{K-1}{t}+(K-1)\binom{K-2}{t-1}+(K-S-1)\binom{K-2}{t}\label{eq:redn}\\&=\binom{K-1}{t}+S\binom{K-2}{t-1}\notag\\&+(K-S-1)\left(\binom{K-2}{t-1}+\binom{K-2}{t}\right)\notag\\&=\binom{K-1}{t}+S\binom{K-2}{t-1}+(K-S-1)\binom{K-1}{t}\notag\\&=(K-S)\binom{K-1}{t}+S\binom{K-2}{t-1}=Q\notag
		\end{align}
	\endgroup}
		coded subfiles of \(W_{d_k}\). Thus, the user \(U_k\) can decode \(W_{d_k}\). This completes the proof of Theorem \ref{thm_pcd2d}. Schemes at other memory points can be obtained by memory sharing.
\end{IEEEproof}
\begin{rem}
	It is possible to further improve the load-memory trade-off in Theorem \ref{thm_pcd2d} by not making certain transmissions in the delivery phase. This reduction is enabled by observing the inequality in \eqref{eq:redn}. From the delivery phase described in the scheme in Theorem \ref{thm_pcd2d}, note that each selfish user receives \(\binom{K-2}{t}\) extra coded subfiles than it requires to decode the demanded file. Also, each non-selfish user makes \(\binom{S}{t+1}\) transmissions solely benefiting the set of selfish users. Of the total \((K-S)\binom{S}{t+1}\) such transmissions, even if any \(\binom{K-2}{t}\) transmissions are not made, the users can recover their demanded files. Therefore, with cache memory
	$M=\frac{N(t+1)(K-1)}{(K-S)(K-1)+tS}, $
	the load
	$R(M) = \frac{(K-S)\binom{K-1}{t+1}-\min\left((K-S)\binom{S}{t+1},\binom{K-2}{t}\right)}{(K-S)\binom{K-1}{t}+S\binom{K-2}{t-1}}$
	is achievable.

	Furthermore, the case $t=0$ corresponds to the pair $(M = N/(K-S),\, R = K-1)$. 
	However, a smaller load of $R = K - \tfrac{K}{K-S}$ is sufficient to satisfy the demands of all users when $M = N/(K-S)$. 
	This reduction is achieved by omitting the transmission $Y_{d_s,\varnothing}^{(k)}$ for any one $k \in \mathcal{K}$ and for every $s \in [K]\setminus \mathcal{K}$. 
	The reason is that each selfish user already stores one coded subfile of its requested file in its cache, so the transmitting users only need to deliver $(K-S-1)$ coded subfiles to ensure decodability. Therefore, the required number of transmissions is $(K-1)(K-S)-S$. Thus, $R=\frac{(K-S)(K-1)-S}{K-S}= K - \tfrac{K}{K-S}$.
	
	Notice that both load-reduction strategies discussed in this remark require additional coordination among the transmitting users in designing the coded transmissions.
	
\end{rem}
The following example illustrates Theorem \ref{thm_pcd2d} and the corresponding partially cooperative D2D coded caching scheme.
\begin{example}
	Consider a $(6,2,6)$ partially cooperative D2D coded caching network. 
	Assume that $t=2$. Each file \(W_n, n \in [6]\) is divided into \((K-S)\binom{K-1}{t}+S\binom{K-2}{t-1}=48\) non-overlapping subfiles, i.e., $W_n = \left\{W_{n,i}:i\in \left[48\right]\right\}, \forall n \in [6].$ These subfiles are then encoded using a $[60,48]$ MDS code. The coded subfiles of each file $W_n$ is represented as $\left[Y_{n,j}:j\in  \left[60\right]\right]$.

	Every \(j\in [60]\) can be uniquely represented as $j=(k-1)10+\phi_k(\mathcal{T})$ 
	for some \(k\in [6]\) and \(\mathcal{T}\subseteq [6]\backslash\{k\}\) with \(|\mathcal{T}|=2\), where the function \(\phi_k:\binom{[6]\backslash \{k\}}{2}\rightarrow [10]\) maps a subset \(\mathcal{T}\) to its lexicographic index in \(\binom{[6]\backslash \{k\}}{2}\). Then, for every \(n\in [6]\) and \(j\in [60]\), we define $Y_{n,\mathcal{T}}^{(k)}\triangleq Y_{n,j}$. For example, for $j \in [11:20]$, choose $k=2$. Then $j=10+\phi_2(\mathcal{T})$, where $\mathcal{T}$ takes values from the ordered set $\{\{1,3\},\{1,4\},\{1,5\},\{1,6\},\{3,4\},\{3,5\},\{3,6\},\{4,5\},\\ \{4,6\},\{5,6\}\}$. The mapping $\phi_2(\mathcal{T})$ assigns $[1:10]$ to these sets, respectively, in the listed order. Therefore, the coded subfiles $Y_{n,11},Y_{n,12},....,Y_{n,20}$ can be uniquely denoted as $Y_{n,\{1,3\}}^{(2)},Y_{n,\{1,4\}}^{(2)},...,Y_{n,\{5,6\}}^{(2)}$, respectively. 
	
	The cache content of user $U_k, \forall k \in [6]$, denoted by  $Z_k$, is given by 
	$Z_k = \left\{Y_{n,\mathcal{T}}^{(k)}: \mathcal{T}\subseteq[6]\backslash \{k\},|\mathcal{T}|=2,n\in [6]\right\}\cup  \qquad\qquad\left(\bigcup_{\ell \in [6]\backslash \{k\}}\left\{Y_{n,\mathcal{T}}^{(\ell)}: \mathcal{T}\ni k, \mathcal{T}\subseteq[6]\backslash \{\ell\},|\mathcal{T}|=2,n\in [6]\right\}\right).$ For example, $Z_1 = \left\{Y_{n,\mathcal{T}}^{(1)}: \mathcal{T} \in \binom{\{2,3,4,5,6\}}{2},n\in [6]\right\}\cup  \qquad\qquad\left(\displaystyle \bigcup_{\ell \in \{2,3,4,5,6\}}\left\{Y_{n,\mathcal{T}}^{(\ell)}: \mathcal{T}\ni 1, \mathcal{T}\in \binom{[6]\backslash \{\ell\}}{2},n\in [6]\right\}\right)= \left\{Y_{n,\{2,3\}}^{(1)},Y_{n,\{2,4\}}^{(1)},Y_{n,\{2,5\}}^{(1)},Y_{n,\{2,6\}}^{(1)},Y_{n,\{3,4\}}^{(1)},Y_{n,\{3,5\}}^{(1)},\right. \\ \left. Y_{n,\{3,6\}}^{(1)},Y_{n,\{4,5\}}^{(1)},Y_{n,\{4,6\}}^{(1)},Y_{n,\{5,6\}}^{(1)} \right\} \cup \left(\left\{Y_{n,\{1,3\}}^{(2)},Y_{n,\{1,4\}}^{(2)},\right. \right. \\ \left. \left.Y_{n,\{1,5\}}^{(2)}, Y_{n,\{1,6\}}^{(2)}\right\}\cup \left\{Y_{n,\{1,2\}}^{(3)},Y_{n,\{1,4\}}^{(3)},Y_{n,\{1,5\}}^{(3)},Y_{n,\{1,6\}}^{(3)}\right\}  \right. \\  \left.\cup \left\{Y_{n,\{1,2\}}^{(4)},Y_{n,\{1,3\}}^{(4)},Y_{n,\{1,5\}}^{(4)},Y_{n,\{1,6\}}^{(4)}\right\} \cup \left\{Y_{n,\{1,2\}}^{(5)},Y_{n,\{1,3\}}^{(5)},\right. \right. \\ \left. \left.Y_{n,\{1,4\}}^{(5)},Y_{n,\{1,6\}}^{(5)}\right\} \cup \left\{Y_{n,\{1,2\}}^{(6)},Y_{n,\{1,3\}}^{(6)},Y_{n,\{1,4\}}^{(6)},Y_{n,\{1,5\}}^{(6)}\right\} \right)$.
	Note that there are $30$ coded subfiles of each of $6$ files in user $U_1$'s cache, and each coded subfile is of $\frac{1}{48}$ file size. Therefore, $M=6 \times 30 \times \frac{1}{48} = \frac{15}{4}$ files. 
	
	Let the demand vector $\vec{d}=(1,2,3,4,5,6)$. Also, let $U_4$ and $U_5$ be the selfish users. Thus the set of transmitting users is \(\mathcal{U}_{\{1,2,3,6\}}=\{U_1,U_2,U_3,U_6\}\). User $U_k$, \(k\in\{1,2,3,6\}\) makes coded transmissions corresponding to every set \(\mathcal{S} \in \binom{[6]\backslash\{k\}}{3} \). Corresponding to each $\mathcal{S}$, user $U_k$, \(k\in \mathcal{K}\), transmits $\bigoplus_{s\in \mathcal{S}} Y_{d_s,\mathcal{S}\backslash \{s\}}^{(k)}$. Each of the four transmitting users transmit $|\binom{[6]\backslash\{k\}}{3}|=\binom{5}{3}=10$ coded subfiles. Therefore, the transmission load $R=4 \times 10 \times \frac{1}{48} =\frac{5}{6}$.
	
	Now, let us consider user $U_1$ and examine how it gets the demanded file $W_1$. User $U_1$ already has $30$ coded subfiles of file $W_1$ in its cache. It requires $18$ more coded subfiles of $W_1$ to decode $W_1$,  since we have used a $[60,48]$ MDS code to encode the subfiles. Consider a user from $\mathcal{U}_{\{1,2,3,6\}}$ other than $U_1$, say $U_2$. Consider a set \(\mathcal{T}\subseteq [6]\backslash \{1,2\}\) such that \(|\mathcal{T}|=2\), i.e., \(\mathcal{T} \in \{\{3,4\},\{3,5\},\{3,6\},\{4,5\},\{4,6\},\{5,6\}\}\). Corresponding to each $\mathcal{T}$, define the set \(\mathcal{S} = \mathcal{T}\cup\{1\}\), i.e., $\mathcal{S} \in \{\{1,3,4\},\{1,3,5\},\{1,3,6\},\{1,4,5\},\{1,4,6\},\{1,5,6\}\}$. Corresponding to $\mathcal{S}=\{1,3,4\}$, user $U_2$ transmits $Y_{d_1,\{3,4\}}^{(2)} \oplus Y_{d_3,\{1,4\}}^{(2)} \oplus Y_{d_4,\{1,3\}}^{(2)}=Y_{1,\{3,4\}}^{(2)} \oplus Y_{3,\{1,4\}}^{(2)} \oplus Y_{4,\{1,3\}}^{(2)}$. From this transmission, user $U_1$ will get the coded subfiles $Y_{1,\{3,4\}}^{(2)}$ of the demanded file $W_1$ since it have the subfiles $Y_{3,\{1,4\}}^{(2)}$ and $Y_{4,\{1,3\}}^{(2)}$ in its cache. Similarly, from the transmissions by user $U_2$ corresponding to $\mathcal{S} = \{1,3,5\},\{1,3,6\},\{1,4,5\},\{1,4,6\} \text{ and }\{1,5,6\}\}$, user $U_1$ will get the coded subfile $Y_{1,\{3,5\}}^{(2)},Y_{1,\{3,6\}}^{(2)},Y_{1,\{4,5\}}^{(2)},\\Y_{1,\{4,6\}}^{(2)} \text{ and } Y_{1,\{5,6\}}^{(2)} $, respectively. That is, $U_1$ gets $6$ coded subfiles of the demanded file $W_1$ from the transmissions from $U_2$. Similarly, it gets another $6$ coded subfiles from each of $U_3$ and $U_6$. Therefore, from the transmissions, $U_1$ gets a total of $18$ coded subfiles of $W_1$, which was the remaining number of coded subfiles it needed. Thus, $U_1$ can decode the file $W_1$.   
\end{example}

Next, we derive a lower bound on the transmission load of a partially cooperative D2D coded caching scheme using the cut-set argument. The condition $M\geq \frac{N}{K-S}$ is needed to ensure that any possible demands can be met using the cache contents of the $(K-S)$ non-selfish users. The following lower bound is derived under the assumption of a symmetric load, i.e., we assume that the transmissions made by each non-selfish user are of equal size. In other words, the normalized size of the transmissions from user \(U_k\) is \(R_k = R/(K-S)\) for every \(k\in \mathcal{K}\).
\begin{figure*}[!htbp]
	\centering
	\begin{subfigure}[t]{0.325\textwidth}
		\centering
		\includegraphics[width=\linewidth]{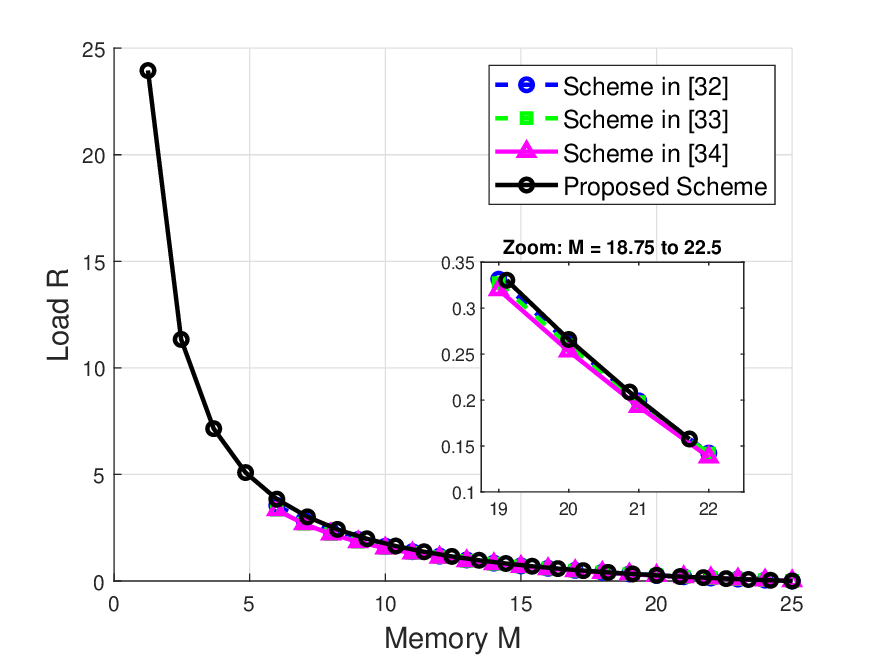}
		\caption{$S=5$}
		\label{fig:sub1}
	\end{subfigure}
	\hfill
	\begin{subfigure}[t]{0.325\textwidth}
		\centering
		\includegraphics[width=\linewidth]{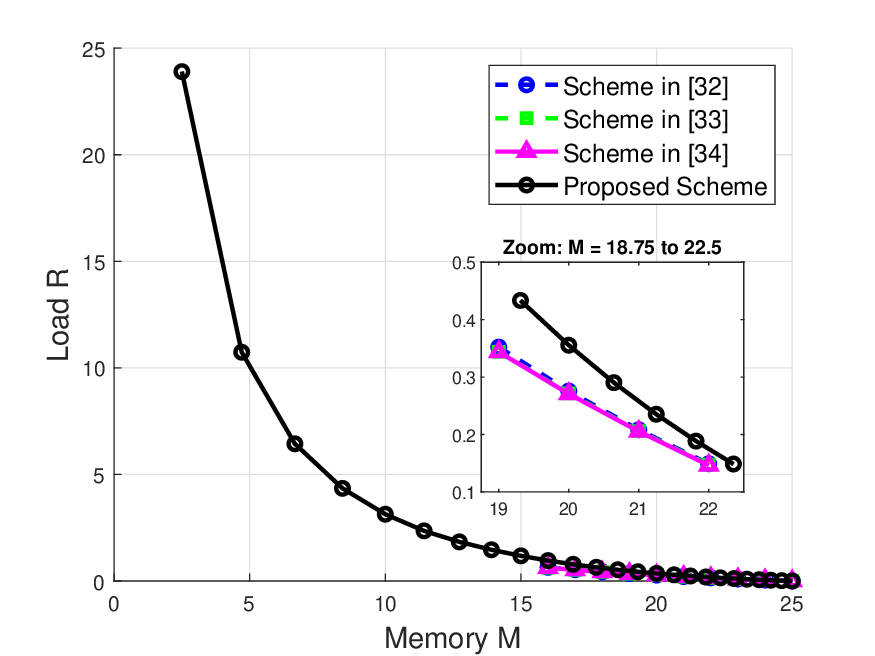}
		\caption{$S=15$}
		\label{fig:sub2}
	\end{subfigure}
	\hfill
	\begin{subfigure}[t]{0.325\textwidth}
		\centering
		\includegraphics[width=\linewidth]{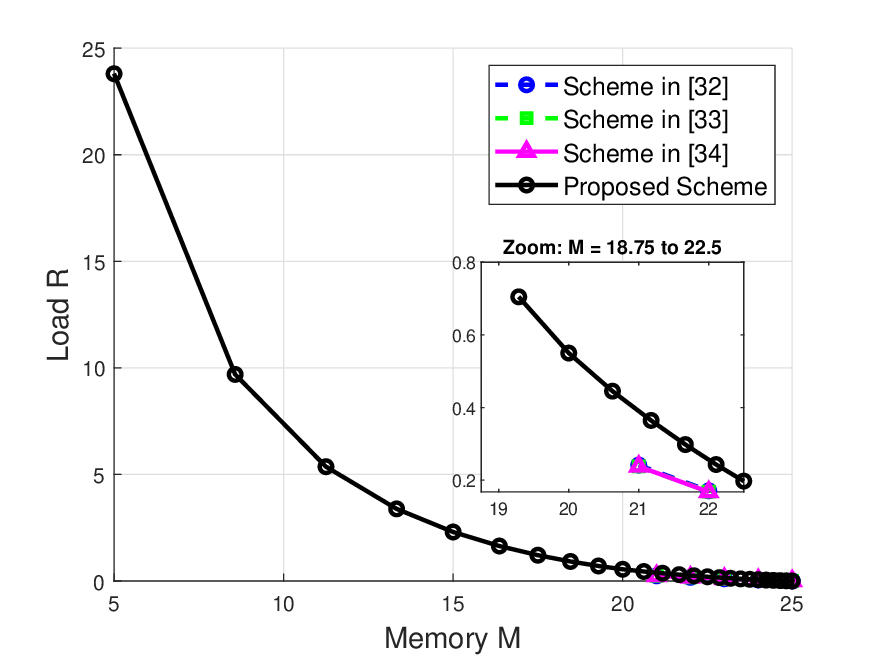}
		\caption{$S=20$}
		\label{fig:sub3}
	\end{subfigure}
	
	\caption{Load–memory trade off of the proposed scheme, the schemes in \cite{TS}, \cite{PV}, and Scheme A in \cite{GHXZCH} for a $(25,S,25)$ partially cooperative D2D network with different values of $S$.}
	\label{fig:pcd2d_comp1}
\end{figure*}
\begin{thm}\label{thm_lb}
	For the \((K,S,N\geq K)\) partially cooperative D2D coded caching scheme, the optimal-load memory trade-off is lower bounded by
		\begin{align}\label{eq:lb}
		R^*(M) &\geq \max_{s\in [K-1]} \frac{N-sM}{\frac{\max(K-S-s,1)}{K-S}\ceil{N/s}}.
		\end{align}
\end{thm}
\begin{IEEEproof}
	Let \(s\in [K-1]\). Consider \(s\) users in the network such that the set of remaining \(K-s\) users includes at least one non-selfish user. Of the considered \(s\) users, assume that \(\ell\leq \min(s,S)\) are selfish users. Since the set of \(s\) users can include at most \(K-S-1\) users, we have \(\ell\geq \max(0,s-(K-S)+1)\). Without loss of generality, we assume that the chosen \(s\) users are indexed from 1 to \(s\). Now, consider the demand vector \(\vec{d}_1=(1,2,\dots,s,\emptyset,\dots,\emptyset)\), where user \(U_k\) demands for the file \(W_k\) for every \(k\in [s]\). The demands of the rest of the users are arbitrary. Let \(\mathcal{U}\) be the set of non-selfish users not part of the set of considered \(s\) users. Note that \(|\mathcal{U}| = (K-S)-(s-\ell)\geq 1\). Let \(\tilde{X}_1\) denotes the transmissions from users in \(\mathcal{U}\) corresponding to the demand vector \(\vec{d}_1\). Now, consider another demand vector \(\vec{d}_2=(s+1,s+2,\dots,2s,\emptyset,\dots,\emptyset)\) and denote the transmission from users in \(\mathcal{U}\) corresponding to \(\vec{d}_2\) as \(\tilde{X}_2\). Similarly, we consider \(\ceil{N/s}\) such demand vectors to cover the entire file library and the corresponding transmissions. Then, we get
	\begin{subequations}
		\begin{align}
			N&=H(W_{[1:N]})\\
			&\leq H(Z_{[1:s]},\tilde{X}_{[1:\ceil{N/s}]})\label{eq:b}\\
			&\leq H(Z_{[1:s]})+H(\tilde{X}_{[1:\ceil{N/s}]})\\
			&\leq sM+\frac{(K-S)-(s-\ell)}{K-S}\left\lceil{\frac{N}{s}}\right\rceil R^*(M),\label{eq:d}
		\end{align}
	\end{subequations}
	where \eqref{eq:b} follows from the fact that the entire file library can be decoded using the cache contents of the chosen \(s\) users and transmissions \(\tilde{X}_{[1:\ceil{N/s}]}\) from users in \(\mathcal{U}\). In addition, \eqref{eq:d} follows from our symmetric load assumption, where the size of the transmissions from users in \(\mathcal{U}\) is \(|\mathcal{U}|/(K-S)\) fraction of the load. Rearranging the final inequality, for every choices of \(s\in [K-1]\) and \(\ell \in [\max(0,s-(K-S)+1):\min(s,S)]\), we have
	\begin{equation}
		R^*(M)\geq \frac{N-sM}{\frac{(K-S)-(s-\ell)}{K-S}\left\lceil{\frac{N}{s}}\right\rceil}.
	\end{equation}
	Since the RHS of the inequality decreases with increasing \(\ell\), in order to maximize the lower bound, we choose \(\ell = \max(0,s-(K-S)+1)\). Therefore, we get
	\begin{align*}
		R^*(M) &\geq \max_{s\in [K-1]} \frac{N-sM}{\frac{(K-S)-(s-\max(0,s-(K-S)+1))}{K-S}\ceil{N/s}}\notag\\ &=\max_{s\in [K-1]} \frac{N-sM}{\frac{\max(K-S-s,1)}{K-S}\ceil{N/s}}.
	\end{align*}
	This completes the proof of Theorem \ref{thm_lb}.
\end{IEEEproof}

\section{Performance Analysis}\label{perform_anlysis_pcd2d}
In this section, performance analysis of the proposed scheme in Section \ref{scheme_pcd2d} is carried out by comparing with all known partially cooperative D2D coded caching schemes, and also by showing the optimality of the proposed scheme in a certain memory regime.
\setcounter{figure}{3}
\begin{figure*}[b]
	\centering
	\begin{subfigure}[!htbp]{0.325\textwidth}
		\centering
		\includegraphics[width=\linewidth]{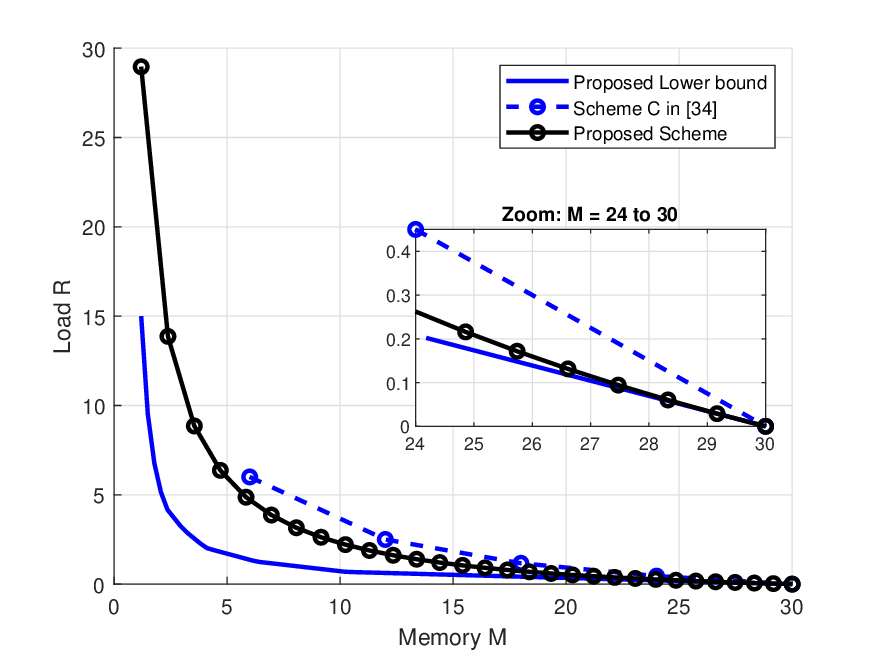}
		\caption{$S=5$}
	\end{subfigure}
	\hfill
	\begin{subfigure}[!htbp]{0.325\textwidth}
		\centering
		\includegraphics[width=\linewidth]{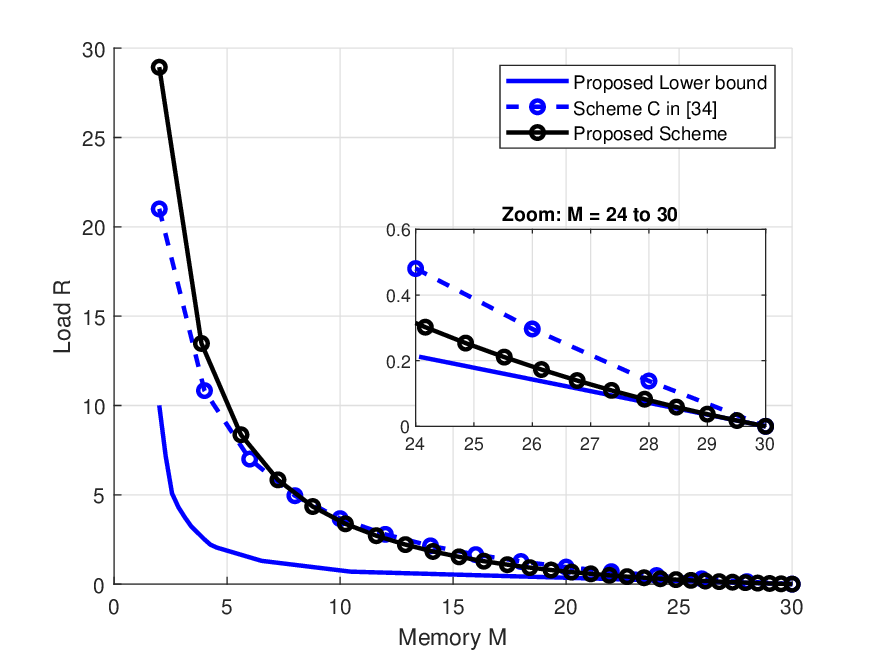}
		\caption{$S=15$}
	\end{subfigure}
	\hfill
	\begin{subfigure}[!htbp]{0.325\textwidth}
		\centering
		\includegraphics[width=\linewidth]{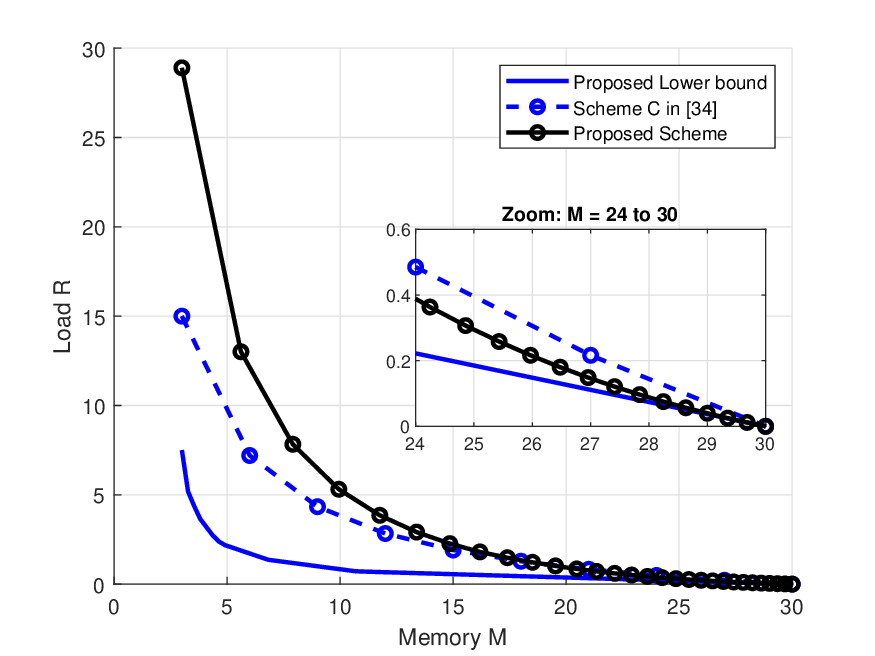}
		\caption{$S=20$}
	\end{subfigure}
	
	\caption{Load–memory trade off of the proposed scheme and Scheme C in \cite{GHXZCH} for a $(30,S,30)$ partially cooperative D2D network with different values of $S$.}
	\label{fig:SchmC_comp1}
\end{figure*}
Coded caching schemes for a $(K, S, N)$ partially cooperative D2D network were studied so far only in \cite{TS, PV, GHXZCH}. However, the deterministic scheme in \cite{TS}, the scheme in \cite{PV} and the Scheme A in \cite{GHXZCH}  operate in higher memory regimes, that is when $M \ge \frac{N}{K}(S+1)$, and also need to know the identity of all the selfish users to design the transmissions in the delivery phase. Scheme B in \cite{GHXZCH} also operates when $M \ge \frac{N}{K}(S+1)$ and requires knowledge of the identities of all selfish users before the placement phase. Moreover, Scheme B in \cite{GHXZCH} was discussed only for $S=1$.  Compared to these schemes, our proposed scheme in Section \ref{scheme_pcd2d} additionally operates in the memory regime $\frac{N}{K-S} \le M \le \frac{N}{K}(S+1)$. That is, the proposed scheme operates in all feasible memory regimes. Unlike these schemes, the proposed scheme does not require knowledge of the identity of selfish users to design the placement phase or delivery phase; it only needs to know the number of selfish users in the D2D network. A comparison of the above-mentioned schemes and the proposed scheme for a $(25, S,25)$ partially cooperative D2D network, for different values of $S$, is shown in Fig.\ref{fig:pcd2d_comp1}, which clearly shows the memory regimes where only the proposed scheme works. It is observed that, in a certain memory regime where all schemes work, the proposed scheme has a higher transmission load compared to others. However, its subpacketization level is lower, as evident from Fig.\ref{Fig:subpack_comp}.
\setcounter{figure}{2}
\begin{figure}[H]
	\centering
	\includegraphics[width=0.9\linewidth]{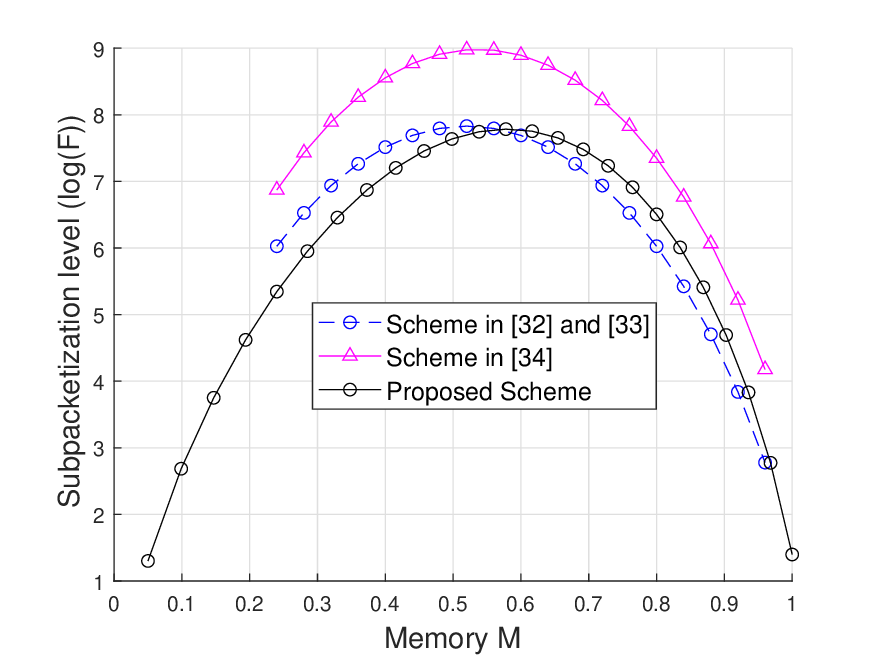}
	\caption{Subpacketization level–memory trade off of various schemes for a $(25,5,25)$ partially cooperative D2D network}
	\label{Fig:subpack_comp}
\end{figure}

Scheme C in \cite{GHXZCH} operates in lower memory regimes also, but requires knowledge of the identity of selfish users before the placement phase, which is often impractical. However, we carry out a comparison with this scheme. Compared to the Scheme C in \cite{GHXZCH}, the proposed scheme achieves a lower transmission load in the entire memory regime when the number of selfish users $S$ is small, and in the higher memory regime for all $S$. This is illustrated with examples in Fig.\ref{fig:SchmC_comp1}.  

Next, we show the optimality of the proposed scheme in a certain memory regime  with respect to the lower bound in Theorem \ref{thm_lb}, as stated in the following theorem. 
\begin{thm}
	\label{thm:opt}
	The proposed scheme is optimal when \( \frac{1}{1+\frac{K-S-1}{(K-1)^2}}\leq \frac{M}{N}\leq 1\) and the corresponding optimal load is 
	\begin{equation}
		R^*(M)=\frac{K-S}{K-S-1}\left(1-\frac{M}{N}\right).
	\end{equation}
\end{thm}
\begin{IEEEproof}
	By substituting, \(s=1\) in \eqref{eq:lb}, we have the following lower bound on the optimal memory-load trade-off
	\begin{equation*}
		R^*(M)\geq\frac{K-S}{K-S-1}\left(1-\frac{M}{N}\right).
	\end{equation*}
	By substituting \(t=K-2\) in \eqref{eq:thm_pcd2d}, we get 
	\(M_1=\frac{N}{1+\frac{K-S-1}{(K-1)^2}}\) and the corresponding load achieved is 
	\(R_1=\frac{K-S}{K(K-1)-S}\). Similarly, corresponding to \(t=K-1\), the memory-load pair \((M_2,R_2)=(N,0)\) is achievable. By memory sharing, the load 
	\begin{equation*}
		R(M)=\frac{K-S}{K-S-1}\left(1-\frac{M}{N}\right)
	\end{equation*}
	is achievable for every \( \frac{1}{1+\frac{K-S-1}{(K-1)^2}}\leq \frac{M}{N}\leq 1\). Therefore, we have the optimal load
	\begin{equation*}
		R^*(M)=\frac{K-S}{K-S-1}\left(1-\frac{M}{N}\right)
	\end{equation*}
	for every \( \frac{1}{1+\frac{K-S-1}{(K-1)^2}}\leq \frac{M}{N}\leq 1\). This completes the proof of Theorem \ref{thm:opt}.
\end{IEEEproof}
The proposed lower bound is plotted in examples in Fig.\ref{fig:SchmC_comp1}. 
In \cite{CKRG}, the authors considered D2D coded caching with user inactivity, modeling each user as being active with some non-zero probability. In their scheme, all active users participate in the delivery phase, and no distinction is made between selfish and non-selfish users. In contrast, in our approach, only a subset of users transmits, while decodability must be ensured for the entire set of users. As a result, a direct comparison with the memory–load trade-off of the scheme in \cite{CKRG} is not directly applicable and could be misleading.

\section{Conclusion}\label{concl_pcd2d}
In this work, we first proposed a novel partially cooperative D2D coded caching scheme. Unlike existing schemes, the proposed scheme does not require knowledge of the identity of selfish users to design the placement phase or delivery phase; it only needs to know the number of selfish users in the D2D network. Compared with existing schemes, the proposed scheme operates in new memory regimes. In the regime where other schemes also operate, the proposed scheme has an advantage in either subpacketization level or transmission load. A lower bound on the load of a partially cooperative D2D coded caching scheme is also obtained. Using this bound, the proposed scheme is shown to be optimal in the high-memory regime.
\section*{Acknowledgement}
This work was supported partly by the Science and Engineering Research Board (SERB) of Department of Science and Technology (DST), Government of India, through J.C Bose National Fellowship to B. Sundar Rajan.	

\end{document}